\documentclass[prl,aps,twocolumn,floats,showpacs]{revtex4}

\usepackage{graphics}
\usepackage{amsmath}
\usepackage{citesort}
\usepackage{dcolumn}

\begin{document}

\title{
{\rm\small\hfill (Phys. Rev. Lett., accepted)}\\
Effect of surface nanostructure on temperature programmed reaction spectroscopy:\\
First-principles kinetic Monte Carlo simulations of CO oxidation at RuO$_2$(110)}

\author{Michael Rieger, Jutta Rogal, and Karsten Reuter}

\affiliation{Fritz-Haber-Institut der Max-Planck-Gesellschaft,
Faradayweg 4-6, D-14195 Berlin, Germany}

\received{27th July 2007}

\begin{abstract}
Using the catalytic CO oxidation at RuO$_2$(110) as a showcase, we employ first-principles kinetic Monte Carlo simulations to illustrate the intricate effects on temperature programmed reaction spectroscopy data brought about by the mere correlations between the locations of the active sites at a nanostructured surface. Even in the absence of lateral interactions, this nanostructure alone can cause inhomogeneities that cannot be grasped by prevalent mean-field data analysis procedures, which thus lead to wrong conclusions on the reactivity of the different surface species.
\end{abstract}


% 68.43.Bc  Ab initio calculations of adsorbate structure and reactions
% 68.43.De  Statistical mechanics of adsorbates
% 68.43.Vx  Thermal desorption
% 82.65.+r  Heterogeneous catalysis at surfaces, surf. chemistry

\maketitle

Temperature programmed desorption (TPD) and reaction (TPR) spectroscopy is a frequently employed tool that provides unparalleled insight into the binding energetics of adsorbates or reactants at solid surfaces by recording the amount of desorbing species while ramping the substrate temperature \cite{woodruff94}. The obtained experimental data does contain information about the spatial arrangement of the surface species before and during the ramp, but only in an indirect way. Prevalent experimental data analysis procedures neglect such dependencies and rely almost entirely on a mean-field approach, arriving at expressions considering only the averaged surface coverages $\theta_i$ of species $i$ \cite{woodruff94}. Obvious breakdowns of the mean-field assumption come from strong lateral interactions among the adsorbates or extrinsic defect sites. The intricate consequences on TPD and TPR spectra are well documented in the literature \cite{kang95} and have made surface scientists broadly aware of the limitations of the prevalent mean-field data analysis procedures. With the explicit lattice structure not taken into account, another fundamental source of deficiency of mean-field theories are the neglected correlations between the locations of the active surface sites themselves, i.e. that the mere way how the active sites are arranged on the surface is not accounted for. Within the general move towards the atomic-scale characterization of ever more complex and nanostructured surfaces exhibiting a wide variety of different surface sites this will become an increasingly important factor also for the analysis of TPD and TPR spectra. We illustrate this here for the catalytic CO oxidation at RuO$_2$(110), which is a showcase for a system with only very small lateral interactions. Without appreciable thermodynamic driving force for segregation (and thus for inhomogeneities in the adlayer) one would generally expect a mean-field picture to be valid. Instead we demonstrate that the neglect of the essentially one-dimensional trench like arrangement of the most active surface sites in a mean-field analysis of a set of experimental TPR data \cite{wendt04} for this system leads to qualitatively wrong conclusions about the reactivity of the various surface species.

\begin{figure}
\scalebox{0.53}{\includegraphics{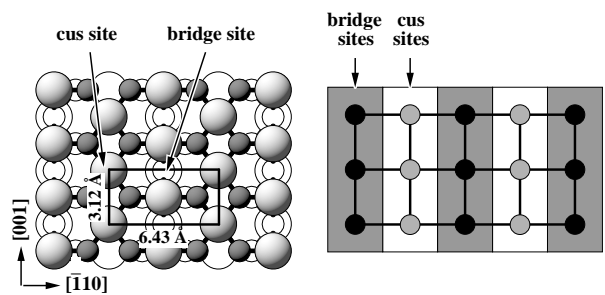}}
\caption{\label{fig1}
Top view of the RuO$_2$(110) surface showing the two prominent active sites (bridge and cus). Ru = light, large spheres, O = dark small spheres. When focusing on these two site types, the surface can be coarse-grained to the lattice model on the right, composed of alternating rows of bridge and cus sites. Atoms lying in deeper layers have been whitened in the top view for clarity.}
\end{figure}

Over the past years, RuO$_2$(110) and the CO oxidation at this model catalyst surface have been extensively studied both experimentally, as well as by first-principles theory \cite{over03}. A wealth of data from ultrahigh vacuum experiments and density-functional theory (DFT) calculations is complemented by detailed {\em in situ} measurements and first-principles kinetic Monte Carlo (kMC) simulations of the steady-state catalytic activity. The picture that is unanimously obtained is that the surface kinetics is predominantly taking place at two prominent active sites, the so-called coordinatively unsaturated (cus) and the bridge (br) site. At the rutile (110) surface these two sites are arranged in alternating rows as illustrated in Fig. \ref{fig1}. Of the corresponding two potentially active oxygen species, O$^{\rm cus}$ and O$^{\rm br}$, it is particularly the much weaker bound (by $\sim 1.5$\,eV) cus oxygen that would be expected to be mostly responsible for the high catalytic activity of this model catalyst at high pressures. This general expectation derives from the established importance of the oxygen-metal bond breaking step in catalytic cycles and the notion behind the well-known Br{\o}nsted-Evans-Polanyi type relationships \cite{bronsted28}. In recent detailed first-principles statistical mechanics modeling this general expectation has been fully confirmed \cite{reuter04}: The calculated O$^{\rm cus}$+CO$^{\rm cus}$ barrier is with 0.9\,eV significantly lower than the one of the competing O$^{\rm br}$ + CO$^{\rm cus}$ reaction (1.2\,eV), and the prior process involving the O$^{\rm cus}$ species was indeed found to completely dominate the simulated steady-state activity of this surface at technologically relevant gas phase conditions. A recent set of TPR experiments came therefore as a particular surprise, since the observed dependence of the CO$_2$ yield on surface coverage was put forward as evidence for a comparable activity of the strongly-bound O$^{\rm br}$ surface oxygen species \cite{wendt04}, thereby implying either a violation of the general Br{\o}nsted-Evans-Polanyi relationships or errors in the relative first-principles barriers of the competing reactions that exceed 0.3\,eV (and thereby amount to several orders of magnitude in the temperature-dependent rate constants).

\begin{figure}
\scalebox{0.85}{\includegraphics{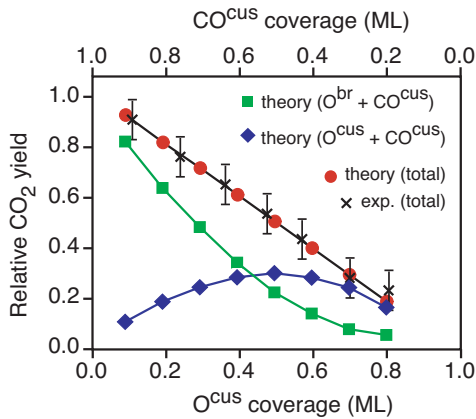}}
\caption{\label{fig2}
(Color online) Total CO$_2$ yield from surfaces initially prepared with varying O$^{\rm cus}$ coverage $\theta$. In all surfaces all bridge sites are initially covered with O$^{\rm br}$ species, and the remaining $(1-\theta)$ cus sites not covered by O$^{\rm cus}$ are occupied by CO$^{\rm cus}$. The CO$_2$ yield is given relative to the one obtained for the surface with zero O$^{\rm cus}$ coverage. Shown are the total simulated CO$_2$ yield, as well as the contributions from the two dominant reaction mechanisms under these conditions: O$^{\rm cus}$+CO$^{\rm cus}$ and O$^{\rm br}$+CO$^{\rm cus}$. The experimental data is taken from ref. \cite{wendt04}.}
\end{figure}

Specifically, TPR spectra were recorded from surfaces in which initially all bridge sites were always fully covered with O$^{\rm br}$, while the coverage of O$^{\rm cus}$ varied in the range $0 < \theta < 0.8$\,monolayer (ML), where 1 ML corresponds to an occupation of all cus sites. The remaining free cus sites were then each time saturated with CO, so that the initially prepared surfaces contained an amount of 1ML O$^{\rm br}$, $\theta$ ML O$^{\rm cus}$ and $(1-\theta)$ ML CO$^{\rm cus}$. Within the understanding of a much more reactive O$^{\rm cus}$ species and thus a dominance of the O$^{\rm cus}$+CO$^{\rm cus}$ reaction, the anticipated coverage dependence of the total CO$_2$ yield in the TPR experiments in a mean-field picture is simply a parabolic $\theta(1-\theta)$. What was measured instead is an almost linear $(1-\theta)$ decrease of the total CO$_2$ yield with increasing O$^{\rm cus}$ coverage as shown in Fig. \ref{fig2} \cite{wendt04}. Such a behavior is much easier rationalized assuming the O$^{\rm br}$ species to be more reactive (and correspondingly the O$^{\rm br}$+CO$^{\rm cus}$ reaction to be dominant), since within the mean-field picture reaction of the constant O$^{\rm br}$ population with the linearly decreasing amount of CO$^{\rm cus}$ at the surface would give rise to a linear $(1-\theta)$ dependence. With the surface kinetics of RuO$_2$(110) dominated by ideal terrace sites and lateral interactions between O and/or CO at these terraces rather small \cite{reuter04}, one would generally expect this mean-field analysis to be applicable, and - together with further experiments on which we comment below - the higher reactivity of the O$^{\rm cus}$ species was correspondingly questioned \cite{wendt04}.

We scrutinize this conclusion with first-principles kMC simulations of the presented TPR data. This technique \cite{kang95,reuter04} combines the accurate description of the individual elementary processes and their interplay with an explicit account of the detailed spatial arrangement of the chemicals at the surface. Here, we consider all non-correlated site and element specific desorption, diffusion and reaction events on the lattice spanned by br and cus sites, employing for each process first-principles rate constants computed by DFT and harmonic transition state theory \cite{reuter04}. This is thus exactly the same setup which was described in detail before and with which the experimentally measured steady-state activity of this surface could be successfully reproduced \cite{reuter04}. The TPR kMC simulations were carried out on $(40 \times 40)$ lattices (800 br, 800 cus sites) with periodic boundary conditions, using the experimental heating ramp of 4.5 K/s. Modeling the dissociative adsorption of O$_2$, surfaces with a defined initial content of O$^{\rm cus}$ species were prepared by randomly filling pairs of neighboring free cus sites until the desired coverage was reached, and then filling all remaining empty cus sites with CO molecules, as well as filling also all bridge sites with O atoms. For such defined initial populations, the TPR spectra were simulated by measuring the amount of desorbing CO$_2$ resolved in 10 K bins during the temperature ramp and averaging over several simulations with different random number seeds. Extensive tests verified that none of the specifics of the numerical setup, and in particular the finite lattice size, affected the quantities reported here. For the discussed total CO$_2$ yields (integrated over the experimentally employed temperature range of 170 - 600 K) slightly wrong desorption peak temperatures obtained as a consequence of the employed DFT energetics are also of lower concern, and we arrive at e.g. essentially identical results and conclusions when employing the refined DFT barriers reported in Ref. \cite{kiejna06}.

Performing our TPR kMC simulations for surfaces prepared with varying cus-site populations, we obtain exactly the same variation of the total CO$_2$ yield with O$^{\rm cus}$ coverage as in the experiments, cf. Fig. \ref{fig2}. The linear dependence interpreted as evidence for a comparable reactivity of the O$^{\rm br}$ species thus results despite the fact that the O$^{\rm cus}$+CO$^{\rm cus}$ reaction with a calculated barrier of 0.9 eV \cite{reuter04} has a temperature-dependent rate constant in our simulations that is always orders of magnitude higher than the one of the competing O$^{\rm br}$ + CO$^{\rm cus}$ reaction with a barrier of 1.2 eV \cite{reuter04}. Disentangling how much of this CO$_2$ yield is produced by which of these two reaction mechanisms in Fig. \ref{fig2}, we see that the linear coverage dependence results in fact as the sum of a parabolic $\theta(1-\theta)$ contribution due to O$^{\rm cus}$+CO$^{\rm cus}$ and a not quite linear $(1-\theta)$ contribution due to O$^{\rm br}$ + CO$^{\rm cus}$. Rather than the qualitative functional form it is thus primarily the relative weight of the two mechanisms that largely differs from the one deduced from the mean-field analysis. 

\begin{figure}
\scalebox{0.24}{\includegraphics{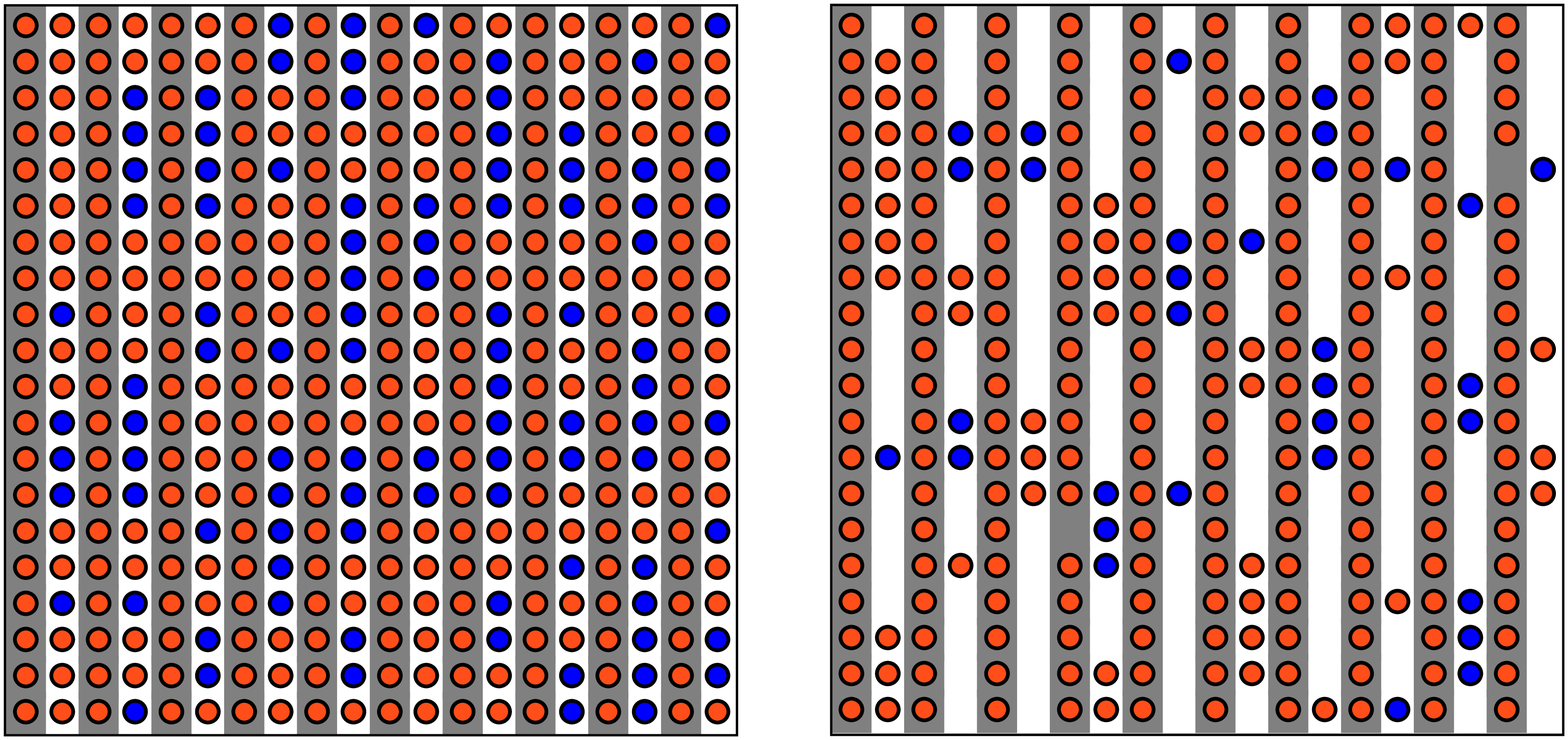}}
\caption{\label{fig3}
(Color online) Snapshots of the surface population in the first-principles TPR kMC simulation for the surface covered initially with 1\,ML O$^{\rm br}$, 0.5\,ML O$^{\rm cus}$ and 0.5\,ML CO$^{\rm cus}$. Shown is a schematic top view of one quarter of the simulation area of $(40 \times 40)$ surface sites, where the substrate bridge sites are marked by gray stripes and the cus sites by white stripes. Oxygen adatoms are drawn as lighter gray (red) circles and adsorbed CO molecules as darker gray (blue) circles. Left panel: Initial population at $T = 170$\,K, Right panel: Population at $T = 400$\,K. At this temperature, according to mean-field $>98$\,\% of all initially present CO$^{\rm cus}$ species should already have been reacted off by the low-barrier O$^{\rm cus}$+CO$^{\rm cus}$ reaction. Instead we find 40\% of them to be still present, namely essentially all those that were not initially adsorbed immediately adjacent to a O$^{\rm cus}$ species.}
\end{figure}

This is most apparent for the surface initially prepared with 0.5 ML O$^{\rm cus}$ and 0.5 ML CO$^{\rm cus}$, where mean-field would predict that essentially all CO$^{\rm cus}$ would be reacted off by the low-barrier O$^{\rm cus}$+CO$^{\rm cus}$ reaction before the O$^{\rm br}$+CO$^{\rm cus}$ reaction can efficiently set in at higher temperatures. Instead we obtain an almost equal contribution of both reaction mechanisms to the total CO$_2$ yield, cf. Fig. \ref{fig2}, which shows that the perfect mixing assumption behind mean-field is not fulfilled at this surface. Intriguingly, the corresponding inhomogeneity in the surface adlayer is here not created by lateral interactions (which are absent in our first-principles model \cite{reuter04}), but follows from the spatial arrangement of the active surface sites and diffusion limitations of the reactants. At the conditions of the TPR experiments the rows of strongly-bound O$^{\rm br}$ species confine the reactive O$^{\rm cus}$ species to one-dimensional cus site trenches, cf. Fig. \ref{fig1}. With the concomitant high diffusion barriers inside these trenches \cite{reuter04}, only a fraction of the CO$^{\rm cus}$ molecules can access the adsorbed O$^{\rm cus}$ atoms and react readily. This is illustrated in Fig. \ref{fig3} which shows snapshots of the surface population at the beginning of the TPR ramp and at a temperature of $T = 400$\,K. This temperature lies above the desorption peak temperature for the O$^{\rm cus}$+CO$^{\rm cus}$ reaction as predicted by mean-field for a second order process and our temperature-dependent first-principles rate constant. According to mean-field essentially all of the initially present CO$^{\rm cus}$ species ($> 98$\,\%) should therefore have already been reacted off. In our kMC simulations we find instead still 40\,\% of them to be present at the surface, namely essentially all those that did not have an immediately neighboring O$^{\rm cus}$ atom in the initially prepared ensemble, cf. Fig. \ref{fig3}. Due to the spatial arrangement of the two prominent active sites, all adsorbed CO$^{\rm cus}$ molecules are on the contrary always chaperoned by two adjacent O$^{\rm br}$ neighbors, enabling a significant fraction of O$^{\rm br}$+CO$^{\rm cus}$ reactions despite the much higher reaction barrier. In this situation it is thus more the neighbor coordination, not the reactivity that determines the relative weight of the two reaction mechanisms in the final CO$_2$ yield, and this cannot be captured within a mean-field approach. Obviously, this geometric effect is specific to the TPR experiments, since under steady-state conditions it would be efficiently counteracted by the replenishment of the cus population due to on-going dissociative O$_2$ and unimolecular CO adsorption. We note, however, that even then a mean-field analysis fails qualitatively, predominantly due to its inability to properly describe the site correlations introduced by the necessity to dissociatively adsorb O$_2$ into two neighboring vacant surface sites \cite{temel07}.

When employing isotope labeling, the contributions of $^{18}{\rm O}^{\rm cus}$ and $^{16}{\rm O}^{\rm br}$ species to the total CO$_2$ yield can also be separated experimentally. Let us assume now that the corresponding two simulated curves in Fig. \ref{fig2} were the data measured in a ``computer experiment''. A reasoning within the prevalent mean-field picture for TPR data analysis would have left us no alternative, but to conclude from the roughly similar contributions of both reaction mechanisms particularly at $\theta = 0.5$\,ML that the O$^{\rm cus}$+CO$^{\rm cus}$ and O$^{\rm br}$+CO$^{\rm cus}$ reaction had roughly equal rate constants. Already at this model surface of moderate complexity and without any lateral interactions, the mere neglection of the way the two prominent active sites are arranged at the surface alone would have thus led us to deduce rate constants that are several orders of magnitude wrong. Or, in terms of reaction barriers we would not have been able to resolve the 0.3\,eV difference in the barriers of the two mechanisms that was really the input to our ``computer experiment''. The errors introduced by the mean-field assumption to the TPR data analysis are thus highly comparable to those obtained when using mean-field based rate theory to extract kinetic parameters from steady-state catalytic activity measurements \cite{temel07}. Such erroneous kinetic parameters are a source of confusion as exemplified by the conjectured violation of the Br{\o}nsted-Evans-Polanyi type relationships for the present RuO$_2$(110) surface \cite{wendt04}, or - maybe even worse regarding theoretical work - this kind of data would then possibly be used to benchmark the quality of first-principles electronic structure calculations. 

Isotope labeling experiments of the just described type, i.e. separating the contributions of $^{18}{\rm O}^{\rm cus}$ and $^{16}{\rm O}^{\rm br}$ species, have in fact been performed and have indeed shown the surprisingly low relative weight of the O$^{\rm cus}$+CO$^{\rm cus}$ reaction in the total CO$_2$ yield \cite{wendt04}. The relative magnitude and qualitative shape of the measured curve are similar to our simulation data shown in Fig. \ref{fig2}, but its maximum is intriguingly shifted from $\theta=0.5$\,ML to $\sim 0.2$\,ML. We are able to explain this finding as a consequence of an insufficient sample preparation. The experimental data was recorded sequentially starting with the $\theta=0.8$\,ML TPR run. During this run a small amount of $^{18}{\rm O}^{\rm cus}$ can diffuse to bridge vacancies created by the $^{16}{\rm O}^{\rm br}+{\rm CO}^{\rm cus}$ reaction. For the consecutive TPR run at the next lower coverage, the surface was only exposed to 20 Langmuir $^{32}{\rm O}_2$ at 750\,K \cite{wendt04,wendt02}. This achieves the intended annealing of remaining bridge vacancies, but does not remove the already present $^{18}{\rm O}^{\rm br}$ species. The ensuing amount of preexisting $^{18}{\rm O}^{\rm br}$ species at each run at lower coverages of $^{18}{\rm O}^{\rm cus}$ falsifies the obtained isotope-labeled CO$_2$ yield and shifts the maximum of the curve, which we are fully able to reproduce when exactly following the described experimental procedure \cite{wendt02} in our simulations \cite{isotope}.
 
In conclusion, we have shown that the detailed spatial arrangement of the active surface sites has a large effect on a recently reported set of TPR experiments studying the catalytic CO oxidation at RuO$_2$(110). Even in the absence of lateral interactions at this surface, this nanostructure alone leads to inhomogeneities that cannot be grasped by prevalent mean-field data analysis procedures, which thus lead to misleading conclusions on the reactivity of the different surface oxygen species. When instead analyzed within an appropriate first-principles statistical mechanics framework the atomic-scale information provided by the present TPR data for RuO$_2$(110) complies fully with the established picture of the function of this model catalyst surface. This highlights the necessity of more refined data analyses when attempting the atomic-scale characterization of ever more complex surfaces exhibiting an ever wider variety of active sites. Correlations between the locations of these different active sites, i.e. their explicit arrangement at the surface, will then play an increasing role, making site accessibility a more and more important factor for the surface kinetics. The RuO$_2$(110) surface studied here exhibits only two prominent active sites, arranged in simple alternating rows. Yet, the mere neglection of this still quite trivial nanostructure leads mean-field TPR data analysis to extract kinetic parameters that are in error by several orders of magnitude and that do not even reflect the relative reactivity of the different surface species correctly.

The EU is acknowledged for financial support under contract NMP3-CT-2003-505670 (NANO$_2$), and the DFG for support within the priority program SPP1091. We thank Horia Metiu and Matthias Scheffler for fruitful discussions and a critical reading of the manuscript.


\begin{thebibliography}{99}

\bibitem{woodruff94}
D.P. Woodruff and T.A. Delchar, {\em Modern Techniques of Surface Science}, Cambridge Univ. Press, Cambridge (1994).

\bibitem{kang95}
e.g. H.C. Kang and W.H. Weinberg, Chem. Rev. {\bf 95}, 667 (1995).

\bibitem{wendt04}
S. Wendt, M. Knapp, and H. Over, J. Am. Chem. Soc. {\bf 126}, 1537 (2004).

\bibitem{over03}
H. Over and M. Muhler, Prog. Surf. Sci. {\bf 72}, 3 (2003);
K. Reuter, Oil \& Gas Science and Technology - Rev. IFP {\bf 61}, 471 (2006); http://ogst.ifp.fr.

\bibitem{bronsted28}
N. Br{\o}nsted, Chem. Rev. {\bf 5}, 231 (1928); 
M.G. Evans and N.P. Polanyi, Trans. Faraday Soc. {\bf 32}, 1333 (1936).

\bibitem{reuter04}
K. Reuter, D. Frenkel, and M. Scheffler,
Phys. Rev. Lett {\bf 93}, 116105 (2004);
K. Reuter and M. Scheffler,
Phys. Rev. B {\bf 73}, 045433 (2006).

\bibitem{kiejna06}
A. Kiejna {\em et al.},
%, G. Kresse, J. Rogal, A. De Sarkar, K. Reuter, and M. Scheffler, 
Phys. Rev. B {\bf 73}, 035404 (2006).

\bibitem{temel07}
B. Temel {\em et al.}, J. Chem. Phys. {\bf 126}, 204711 (2007).

\bibitem{wendt02}
S. Wendt, Ph.D. thesis, FU Berlin (2002).

\bibitem{isotope}
Insufficient sample preparation is also the presumed reason behind the observed desorption of isotope scrambled $^{34}{\rm O}_2$ after exposing the stoichiometric RuO$_2$(110) surface to $^{36}{\rm O}_2$ \cite{wendt04}. Preceding this experiment the surface was only annealed to 600K \cite{wendt02}. In our simulations we see that this is not sufficient to desorb all O$^{\rm cus}$ species, since due to the severe diffusion limitations within the cus trenches there is always a fraction of isolated and at 600\,K still practically immobile O$^{\rm cus}$ atoms remaining. These atoms can then only desorb at a much higher temperature either via recombination with a neighboring O$^{\rm br}$ species or because diffusion is then facilitated and enables associative desorption of two remaining O$^{\rm cus}$ species. Such remnant $^{16}{\rm O}^{\rm cus}$ species that are inadvertently present after the 600\,K annealing and therefore before saturating the cus sites with isotope-labeled $^{18}{\rm O}^{\rm cus}$ provide a natural explanation for the measured fraction of isotope scrambled $^{34}{\rm O}_2$ in the TPD runs. The originally proposed explanation for these findings \cite{wendt04} was an O$^{\rm cus}$-O$^{\rm br}$ exchange diffusion process, which considering the low defect density at this surface would have to take place quite efficiently even at ideal terrace sites. Within our DFT setup \cite{reuter04} we compute a barrier above 2\,eV for such a process, and can therefore rule out this original proposition.

\end{thebibliography}
\end{document}